\documentclass[runningheads]{llncs}

\usepackage{amsmath,amssymb,amsfonts}

\usepackage{graphicx}
\usepackage{subcaption}

\usepackage{graphicx}

\usepackage{xcolor}
\definecolor{Brown}{rgb}{0.5,0.2,0.1}

\usepackage{orcidlink}

\usepackage{enumitem}
\usepackage{todonotes}

\usepackage{hyperref}
\hypersetup{
    colorlinks=true,
    linkcolor=Brown,
    urlcolor=Brown,
    citecolor=Brown,
    filecolor=black,
    bookmarks=true,
    unicode=true,
    pdftitle={},
    pdfauthor={},
    pdfcreator={}
}

\usepackage{cleveref} 

\usepackage{lipsum}
\usepackage{multicol}
\usepackage{dutchcal}
\usepackage{wrapfig}
\usepackage{mathtools}
\usepackage{array}
\usepackage{threeparttable}
\usepackage{multirow}
\usepackage{float}
\usepackage{makecell}
\usepackage{url}
\usepackage{booktabs}
\usepackage[switch]{lineno}
\usepackage{comment}
\usepackage{xspace}

\newcommand{\procset}{\mathcal{P}}
\newcommand{\caseset}{\mathcal{C}}

\newcommand{\eventset}{\mathcal{E}}
\newcommand{\N}{\mathbb{N}}
\newcommand{\proj}{\pi}

\usepackage{algorithmic}

\def\ourmethod{Know Your Streams }
\def\ourgenerator{Stream of Intent }
\newcommand{\ourgeneratorname}{Stream of Intent\xspace}
\titlerunning{Know Your Streams}

\newcommand{\mypar}[1]{\smallskip\noindent\textbf{#1.}}

\begin{document}

\title{Know Your Streams: On the Conceptualization, Characterization, and Generation of Intentional Event Streams}

\author{
    Andrea Maldonado$^*$\inst{1,4,5}\orcidlink{0009-0009-8978-502X} \and
    Christian Imenkamp$^*$\inst{2}\orcidlink{0009-0007-4295-1268} \and
    Hendrik Reiter\inst{3}\orcidlink{0009-0003-8544-0012} \and
    Thomas Seidl\inst{4,5}\orcidlink{0000-0002-4861-1412} \and
    Wilhelm Hasselbring\inst{3}\orcidlink{0000-0001-6625-4335} \and
    Martin Werner\inst{1}\orcidlink{0000-0002-6951-8022} \and
    Agnes Koschmider\inst{2}\orcidlink{0000-0001-8206-7636}
}

\institute{
    School of Engineering and Design, Technical University of Munich, Germany\\
    \email{\{andrea.maldonado\, martin.werner\}@tum.de}
    \and
        University of Bayreuth, Bayreuth, Germany\\
    \email{\{christian.imenkamp,agnes.koschmider\}@uni-bayreuth.de}
    \and
    Christian-Albrechts-University Kiel, Kiel, Germany\\
    \email{\{hendrik.reiter,hasselbring\}@email.uni-kiel.de}
    \and
    Database Systems and Data Mining, Ludwig Maximilian University of Munich, Germany
    \email{seidl@dbs.ifi.lmu.de}
    \and
    Munich Center for Machine Learning, Munich, Germany
}

\authorrunning{A. Maldonado \& C. Imenkamp et al.}
\maketitle
\def\thefootnote{*}\footnotetext{Equal contribution}\def\thefootnote{\arabic{footnote}}
\setcounter{footnote}{0}

\begin{abstract}
    The shift toward IoT-enabled, sensor-driven systems has transformed how operational data is generated, favoring continuous, real-time event streams (ES) over static event logs. This evolution presents new challenges for Streaming Process Mining (SPM), which must cope with out-of-order events, concurrent activities, incomplete cases, and concept drifts. Yet, the evaluation of SPM algorithms remains rooted in outdated practices, relying on static logs or artificially \textit{streamified} data that fail to reflect the complexities of real-world streams. To address this gap, we first perform a comprehensive review of data stream literature to identify stream characteristics currently not reflected in the SPM community. Next, we use this information to extend the conceptual foundation for ES. Finally, we propose Stream of Intent, a prototype generator to produce ES with specific features. Our evaluation shows excellence in producing reproducible, intentional ES for targeted benchmarking and adaptive algorithm development in SPM.  


\end{abstract}

\section{Introduction}
\label{sec:intro}
The rise of the Internet of Things (IoT), Industry 4.0, and distributed sensor systems has fundamentally changed how operational data is generated. Increasingly, organizations must deal not only with static collections of events, but also with Event Streams (ES) i.e., real-time sequences of events emitted by interconnected components and systems.
Consider a manufacturing setting in which multiple machines, equipped with sensors, continuously send status updates to a central system. Here, events arrive asynchronously, possibly out of order, and with limited knowledge about case completion. Moreover, such environments are subject to concept drifts, for instance, when a new machine is introduced or a production line is reconfigured, the underlying process behavior shifts, potentially invalidating previously learned process models. 
These conditions are typical in industrial environments and pose significant challenges for the practical application of process mining (PM).

Traditional PM assumes that event logs are complete, static, chronologically ordered by case, and collected retrospectively. These assumptions work well for postmortem analysis but do not capture the dynamic, real-time nature of ES environments.
In contrast, Streaming Process Mining (SPM) aims to support real-time analysis by processing events on the fly. Although traditional PM can analyze and explain the reasons behind a factory’s production failure after it has occurred, SPM aims to spot issues early enough to prevent them. However, existing approaches often rely on \textit{streamified} (i.e., incremental replay of static) event logs for evaluation \cite{burattin_uncovering_2023,burattin_online_2015,augusto_split_2019,burattin_heuristics_2012,burattin_control-flow_2014} or use comparable synthetic event logs generated by the CPN Tools~\footnote{ \url{http://cpntools.org}} rather than real-time event streams. 
These approximations lack core properties of realistic ES, such as temporal disorder~\cite{9230157}, indefinite case lifespans, and concurrent activities~\cite{burattin_uncovering_2023}, which undermine their utility for comprehensive development and testing. Furthermore, \cite{10.1145/3328905.3332511} generates Out-of-Order streams by assigning some delayed ingestion times to an existing log.

\begin{figure}[h]
     \centering
     \includegraphics[width=0.7\linewidth]{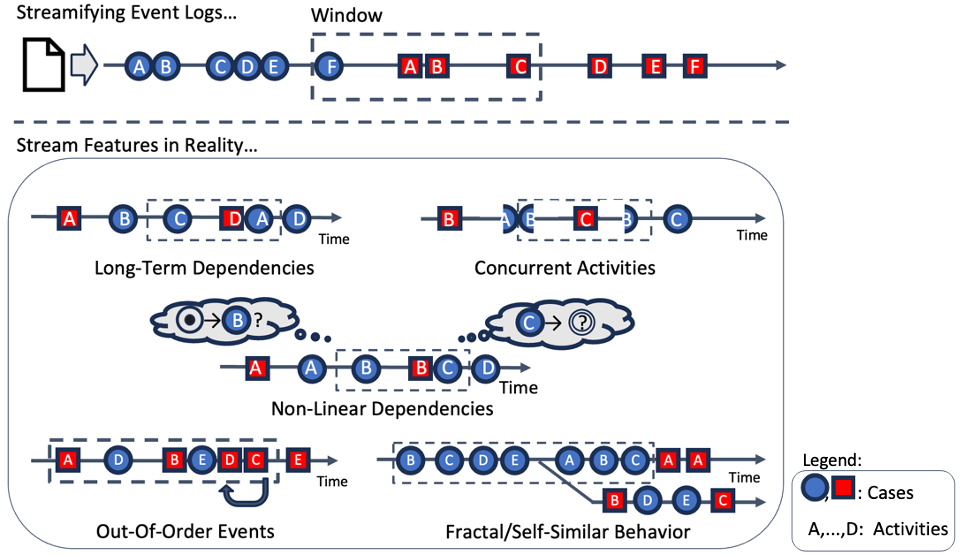}
     \caption{\ourmethod is the first framework to breach ES research with the complexities of real-world real-time processes.}
     \label{fig:motivation}
\end{figure}

To address this gap, we extend the formalization of ES that explicitly captures the distinguishing characteristics of real-world streaming data (\Cref{fig:motivation}). Our structured scheme consolidates previously fragmented definitions of what constitutes an ES and identifies key dimensions that must be considered for evaluation in a streaming context. We conducted this based on a comprehensive literature review of the general data stream community. Building on this basis, we propose a conceptual framework and a prototype generator that produces intentional ES.

In summary, our contributions are as follows: (1) We identify characteristics of ES from the general data stream community and compare them to current SPM approaches. (2) We expand the ES definition and derive a conceptual framework for ES. (3) We propose a prototype to generate ES with intentional characteristics.

\section{The Need for Realistic Event Streams}
\label{sec:motivation_realistic_es}
\begin{figure}[ht]
  \centering
  \begin{subfigure}[t]{0.35\textwidth}
    \vspace{0pt}%
    \centering
    \includegraphics[width=\textwidth]{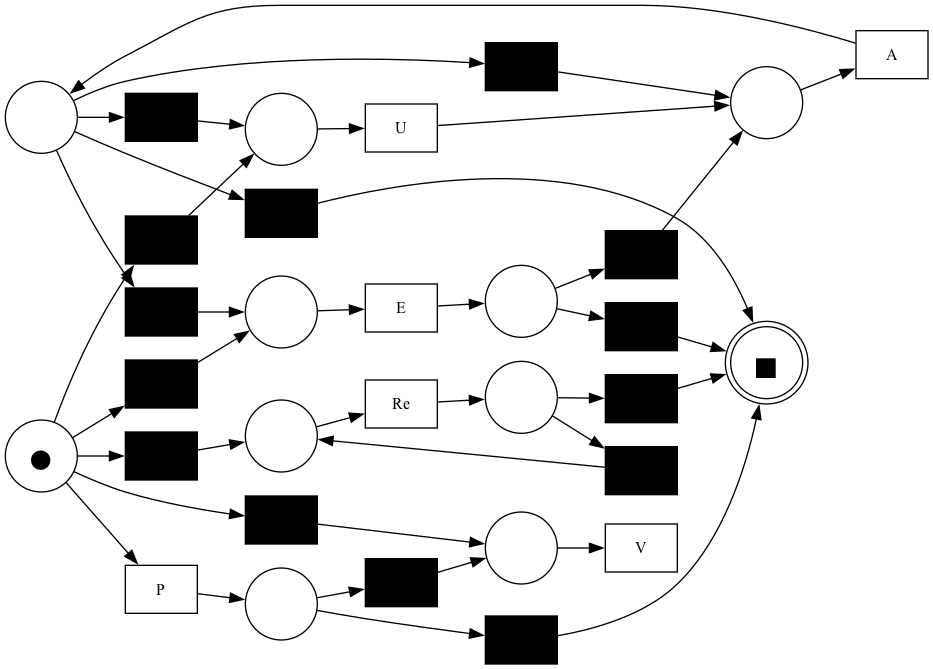}
    \caption{Heuristics miner using a stream-based event log.}
    \label{fig:Baseline_downstream}
  \end{subfigure}
  \hfill
  \begin{subfigure}[t]{0.6\textwidth}
    \vspace{0pt}%
    \centering
    \includegraphics[width=\textwidth]{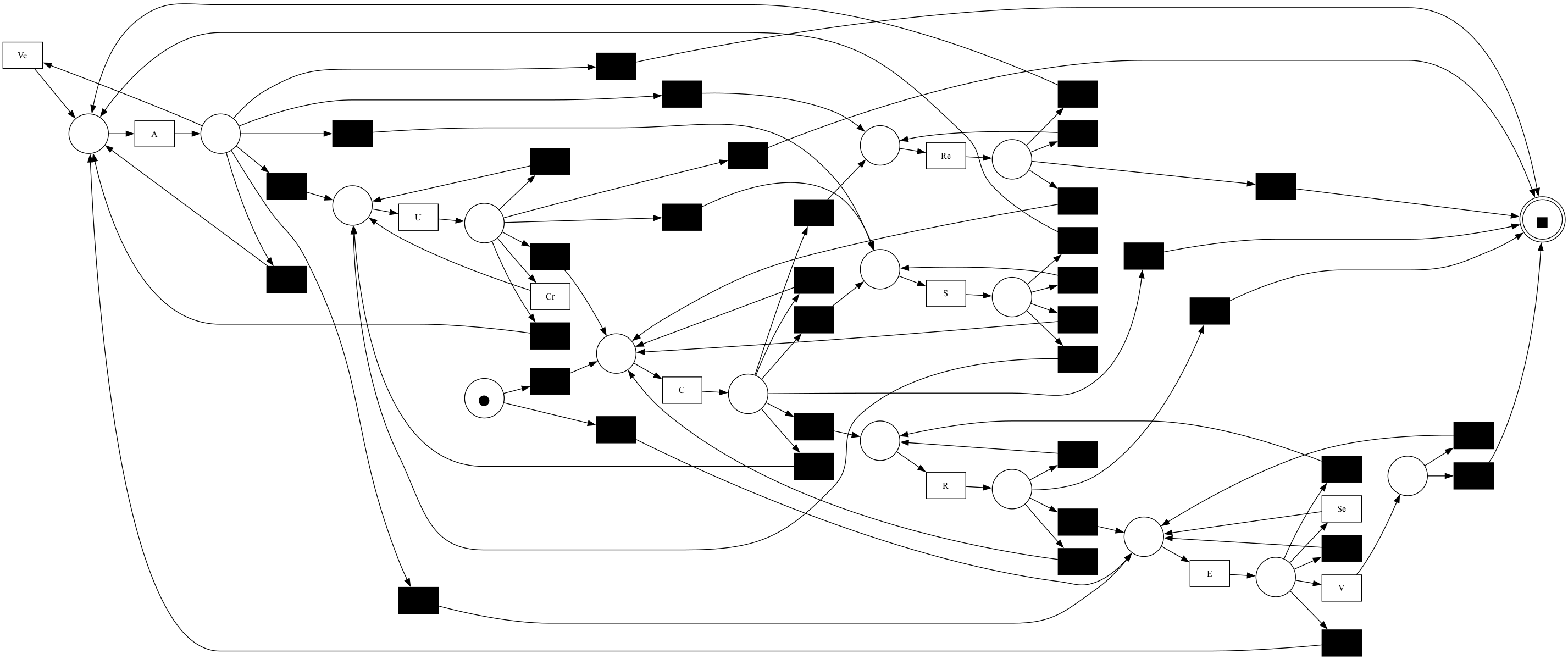}
    \caption{Streaming heuristics miner using an ES containing realistic ES characteristics.}
    \label{fig:Daseline_downstream_broken}
  \end{subfigure}
  \caption{Comparing discovered process models (with $\text{dependency threshold}=0.85$) including vs.\ omitting realistic ES characteristics from the same event source.}
  \label{fig:downstreamtask}
\end{figure}

To motivate the need for realistic ES characteristics, we compare the process discovery results from a stream-based event log, a common approximation, with those from the same ES, which inherently includes the realistic characteristics identified in our work. 
\Cref{fig:downstreamtask}
shows the results. 
Both (a) and (b) required extensive manual preprocessing, a classical step in process mining, to discover the process model (e.g., restoring directly following relations, filtering out fractal behavior).
This is because existing algorithms are not equipped to handle these characteristics.
Consequently, such a time-consuming preprocessing is a prerequisite that is often not feasible in real-time settings.
Instead of relying on extensive filtering and cleaning to convert streams into an "offline" format, we propose integrating realistic stream characteristics directly into algorithms to leverage them for valuable analysis.

In this experiment, we use (a) as to establish a baseline for the expected process model, which was created from a preprocessed version of the stream in (b).
In contrast, (b) presents the result of the streaming heuristics miner \cite{burattin_heuristics_2012}. 
Notably, even the streaming heuristics miner required additional preprocessing to discover the process model in (b), as its design does not inherently handle the structures present in a realistic stream. 
A comparison of the two results reveals a significant increase in silent transitions in ((b) = 36) compared to ((a) = 15), which results in a greater number of required connections.
Furthermore, our analysis shows that the streaming heuristics miner tends to discover more loops when dealing with realistic streams, which is likely a consequence of their more complex and unpredictable nature.

The purpose of this experiment is not to diminish the value of established methods, but to highlight the opportunity to harness the benefits of true real‐time analysis through targeted modeling of realistic stream characteristics.
\section{Related Work}
\label{sec:relatedWork}

Data streams differ significantly from static datasets \cite{aggarwal_framework_2003}. In particular, characteristics of real-time environments (i.e., continuous data arrival, limited resources, latency constraints, and unpredictable future data instances)  pose unique challenges \cite{cormode_references_nodate,chakrabarti_near-optimal_2010,haug_standardized_2022}. These challenges are inherent in the application side of the streaming environment. Most SPM algorithms analyze the data and control-flow of the stream. Given that, we will focus on literature describing characteristics of streams that make it challenging to detect patterns in streaming data. 

Streams include complex patterns and dependencies. For instance, temporal dependencies illustrate the influence of past events on subsequent events \cite{song_learning_2023,cuomo_detecting_2022}. Furthermore, long-term dependencies pose additional challenges, requiring algorithms to maintain and utilize historical information \cite{song_learning_2023}. Non-linear dependencies further increase complexity. Future events often are the result of past non-linear rules rather than simple sequential relationships \cite{chandola_anomaly_2009,aggarwal_framework_2003}.

Moreover, realistic ES frequently experience out-of-order events. This disrupted the timely assumptions often required in traditional methods \cite{song_learning_2023,cormode_references_nodate}. Additionally, fractal behavior, which is marked by repeating patterns across different scales, breaks common assumptions about data being independent and uniformly distributed. This makes tasks such as anomaly detection and forecasting more challenging \cite{zhang_research_2025,de_sousa_measuring_2006}.

However, SPM designs and evaluations are often based on artificially \textit{streamified} event logs and adopt the foundations of classical process mining, ignoring the characteristics of real-time environments \cite{burattin2012heuristics,multi_perspective,10.1007/978-3-031-34560-9_26}. 
\section{Event Streams}
In this section, we consolidate the definition of event stream (ES), particularly focusing on extending the static atomic event notation to handle activities with duration to better capture the complexities of real-world processes and analyze ES characteristics, such as concurrency.
\subsection{Event Stream Formalization}
\label{subsec:event_stream}

Typically, in PM, an \textit{event} $e \in \mathcal{E}: c \times a \times t$ is an atomic tuple in event data.
Each event contains at least three attributes: A \textit{case id} $c \in \mathcal{C}$ from the universe of case identifiers $\mathcal{C}$, an \textit{activity} $a \in \mathcal{A}$ from the universe of activities, $\mathcal{A}$ and a \textit{timestamp} $t \in \mathbb{N}$. 
Further, the projection function $\pi_x$ maps an event to its attribute x. 
Hence, $\pi_c((c_0, a_0, t_0)) = c_0$.

This definition is sufficient to model instantaneous events and many real-world scenarios. These activities naturally require both a start and an end timestamp. Consider a smart factory where machines autonomously perform tasks such as assembling parts, performing diagnostics, or transporting materials. These activities span time intervals: for example, assembling a component may begin at 14:03:10 and complete at 14:04:05. If such activities are represented using a single timestamp, critical temporal information is lost. This simplification limits the kinds of questions to be answered, like the average duration of an activity, the presence of waiting times between activities, or the number of concurrently running cases. These limitations clearly show that an ES model relying on only one timestamp per activity is inadequate for this category of applications.

To address this limitation, we introduce a formalism where each activity is represented by a pair of events: A \textit{start event} marking the beginning of the activity and an \textit{end event} marking its completion.
Let $\mathcal{E}_s$ and $\mathcal{E}_e$ denote the sets of start and end events. We define $e_s=(c,a,t,"start") \in \mathcal{E}_s$ and $e_e=(c,a,t,"end") \in \mathcal{E}_e$. We model the \textit{interval-based} ES as a function:
$S^*:\mathbb{N} \rightarrow \mathcal{E}_s \cup \mathcal{E}_e$. An event stream is called  temporally ordered if the following holds: $\forall i,j \in \mathbb{N}: i < j \implies \pi_t(S^*(i)) \leq \pi_t(S^*(j))$.

To map start events to their respective end events we introduce the function $\mu: \mathcal{E_s} \rightarrow \mathcal{E_e}.$ This function has the following constraints for all $e_s \in \mathcal{E}_s$ and $e_e = \mu(e_s) \in \mathcal{E}_e$: $\pi_c(e_s)= \pi_c(e_e)$, $\pi_a(e_s)= \pi_a(e_e)$ and $\pi_t(e_s) \leq \pi_t(e_e)$.

Building on this formalism, we can now define properties within ES, such as the concurrency of events at a given timestamp $t$ as 
$ \text{concurrency}(t^*) = \left|\{ e_s \in  \mathcal{E}_s\mid \pi_t (e_s) \leq t^* \leq \pi_t (\mu (e_s)) \}\right| $.

Another crucial aspect in real-world ES is the absence of explicit case identifiers \cite{HELAL2022102031}. Events often arrive without a clear reference to a process instance. Therefore, online correlation is an additional key challenge for SPM and stream generation.

\label{sec:formal_framework}

\subsection{Event Stream Characteristics}
\label{subsec:stream_features}
To further build on the holistic definition of an ES, we discuss the characteristics derived from the literature (\Cref{sec:relatedWork}) and their translation from data streams into ES. For that, we use the extended ES formalization shown in \Cref{subsec:event_stream}. In the following section, we propose formalisms for temporal, nonlinear, and long-term dependencies, out-of-order events, and fractal process structures.

\mypar{Event Observation and Temporal Reality}
In real-time environments, it is essential to clearly distinguish between the time an event occurs in the real-world process (the \textbf{event timestamp}) and the time it is captured and received by the analysis system (the \textbf{arrival timestamp}).

We therefore extend the event tuple to $e = (c, a, t, \text{type}, t_{arrival})$, where $t \in \N$ is the event timestamp and $t_{arrival} \in \N$ is the arrival timestamp. The ES, $S^*: \N \rightarrow \eventset_s \cup \eventset_e$, is a sequence ordered by arrival time, reflecting the order of observation.
$$ \forall i, j \in \N: i < j \implies \proj_{t_{arrival}}(S^*(i)) \leq \proj_{t_{arrival}}(S^*(j)) $$
This ordering allows us to formally define events that arrive out of their natural temporal sequence.

\paragraph{Out-of-Order Events}

An ES $S^*$ is considered to contain \textbf{out-of-order events} if the sequence of event timestamps $t$ is not monotonically non-decreasing concerning the arrival sequence. 
Formally, an out-of-order event is any event $e_j = S^*(j)$ in the position $j$ for which a preceding event in the stream, $e_i = S^*(i)$ with a position $i$, where $i < j$, has a later event timestamp. Let $X$ denote the set of relevant attributes used to compare events. An event $e_j = S^{\ast}(j)$ is Out-Of-Order if there exists a preceding event $e_i = S^{\ast}(i)$ with $i < j$ such that $\pi_{t}(e_j) < \pi_{t}(e_i)$ for the same projection on $X$. The displacement $\pi_{t}(e_i) - \pi_{t}(e_j)$ then quantifies the degree of disorder. To avoid ambiguity, we explicitly require that the ordering refers to the attribute $t_{\text{arrival}}$.

$$\exists i, j \in \mathbb{N} \; \big( i < j \;\wedge\; \forall x \in X:\; \proj_t(S^*(j,x)) < \proj_t(S^*(i,x)) \big)$$
The magnitude of this temporal displacement, $\proj_t(S^*(i)) - \proj_t(S^*(j))$, can be used as a measure of the stream's disorder.

\paragraph{Dependencies: Temporal, Long-Term, and Non-Linear}

The evolution of a process case is governed by dependencies between its events. To model this, we define the \textbf{history} of a case $c \in \caseset$ at time $t$, denoted $H(c, t)$, as the ordered sequence of all events belonging to that case up to time $t$.
$$ H(c, t) = \langle e_1, e_2, \dots, e_k \mid \proj_c(e_i) = c \land \proj_t(e_i) \leq t \rangle $$
The generation of subsequent events in the case is determined by a conceptual \textbf{transition function}, $\delta$, which maps the case history to a new activity.
$$ a_{k+1} = \delta(H(c, t)) $$
This function $\delta$ encapsulates the underlying process logic:
\begin{itemize}
    \item \textbf{Temporal Dependencies}: These are inherent, as $\delta$ operates on the event history. The time elapsed between events, $\proj_t(e_{k+1}) - \proj_t(e_k)$, is a direct consequence of the process modeled by $\delta$.
    \item \textbf{Long-Term Dependencies}: These occur when the function $\delta$ is sensitive to events early in the history ($e_i$ for $i \ll k$), meaning distant past events can influence current and future activities.
    \item \textbf{Non-Linear Dependencies}: The complexity of the process logic is embodied by $\delta$. If $\delta$ represents simple sequential logic, the dependencies are linear. However, if $\delta$ models complex branching, parallelism, or context-aware decisions, it gives rise to non-linear dependencies.
\end{itemize}

$\delta$ can be instantiated as a simple mapping where the next activity depends only on the last activity, e.g., $\delta(\langle A \rangle) = B$, or as a context-aware rule that considers the entire history, e.g., $\delta(\langle A, C \rangle) = D$ if an earlier event $C$ occurred. 

\paragraph{Hierarchical Structures and Self-Similarity}

Processes in the real world are often not isolated, but can be nested or can trigger other processes. We can model this with a \textbf{case containment relation}, denoted by the symbol $\prec$. The expression $c_j \prec c_i$ signifies that case $c_j$ is a sub-process or a nested component of case $c_i$. This relation imposes a hierarchical, directed acyclic graph structure on the set of all cases $\caseset$.

A process may exhibit \textbf{fractal or self-similar behavior} if the underlying process logic repeats across different levels of this hierarchy. Let $\procset(c)$ be the abstract process model for a case $c$ (represented by its transition function $\delta$). 

Let $e_k \in c_i$ be the trigger event for case $c_j$. This establishes a temporal correlation in which the existence of $c_j$ depends on $e_k$. The first subprocess event must occur after or at the same time as the triggering event: $\min(\{\pi_t(e) | \pi_c(e) = c_j\}) \ge \pi_t(e_k)$.

Self-similarity exists if, for a pair of cases $c_j \prec c_i$, their process models are structurally related by a transformation $\mathcal{T}$:
$$ \procset(c_j) \approx \mathcal{T}(\procset(c_i)) $$
The transformation $\mathcal{T}$ could represent various relationships, such as a simplification of the parent process, a scaling of its temporal properties, or an abstraction of its activities (as seen in \Cref{fig:motivation}).

$F$ may be realized as a scaling of temporal properties (e.g., multiplying all activity durations by a factor of two) or as a structural simplification (e.g., collapsing a parallel branch into a sequential pattern).




\section{Stream of Intent: Our Prototype Generator}%
\label{sec:method}
ES are fundamental for analyzing and understanding dynamic systems. 
Generating synthetic ES with specific, controllable characteristics is crucial for robust system testing, anomaly detection, and algorithm validation, especially when real-world data is scarce or sensitive. 
Our prototype, \ourgeneratorname, addresses this need by providing a novel approach to create synthetic ES with intentionally predefined features. 
It leverages the underlying simulation capabilities of the Distributed Event Factory (DEF) \cite{reiter_distributed_nodate} and the feature-driven optimization framework of Generating Event Data with Intentional Features (GEDI) \cite{maldonado_gedi_2024}. To define the features required by GEDI we utilize the characteristics presented in \Cref{sec:formal_framework}. We define a feature as a quantifiable characteristic with a numerical value. 

This section details the architecture and method of \ourgeneratorname, explaining how it iteratively optimizes generator parameters to produce streams closely adhering to specified characteristics.

\subsection{Our Approach: Stream of Intent Pipeline}
\label{subsec:approach_overview}
\begin{figure}[b]
    \centering
    \includegraphics[width=\linewidth]{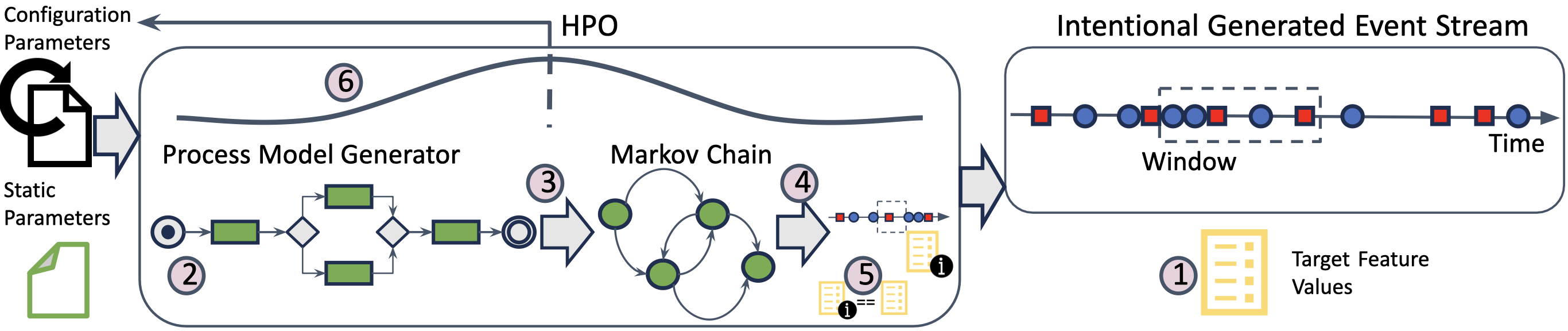}
    \caption{\ourgenerator generates ES with intentional features, optimizing configuration parameters.}
    \label{fig:overview_simple}
\end{figure}
The complete pipeline of \ourgeneratorname{} is conceptually illustrated in Fig.~\ref{fig:overview_simple}.
The generation process begins by defining \textit{static parameters} (e.g., window size, and implicit/explicit time order) and \textit{dynamic configuration parameters} that are subject to optimization.
The core idea is an iterative optimization loop that ensures the generated streams adhere closely to predefined \textit{target feature characteristics}.
The pipeline consists of the following steps:
\begin{enumerate}
    \item \textbf{Specify Stream Features and Targets:} The user defines specific goals for one or more stream features (e.g., target event frequency, specific concurrency patterns). These target values drive the optimization process, similarly to how GEDI defines a target feature vector $g_k$.
    \item \textbf{Generate Process Model:} An initial process tree is generated using stochastic methods, such as those described in \cite{Jouck2016PTandLogGeneratorAG}, along with an initial set of PTLG (Process Tree-based Log Generator) configuration parameters. This process tree serves as a high-level model of the desired process behavior.
    \item \textbf{Process Tree to Markov Chain Conversion:} The generated process tree is converted into a Markov chain, translating its structural and behavioral properties into probabilistic state transitions for DEF's event simulation. The specifics of this conversion are elaborated in Section~\ref{subsec:ptg_to_def}.
    \item \textbf{Generate Stream (DEF Simulation):} The Markov chain derived from the process tree is then used by DEF to simulate a intentional ES. During this simulation, \ourgenerator also accounts for concurrent events, event lifecycles, and event durations, which are critical for realistic stream generation (see Section~\ref{sec:formal_framework}). DEF's underlying Markov chain model, as described in \cite{reiter_distributed_nodate}, allows for the bottom-up generation of events.
    \item \textbf{Feature Extraction and Evaluation:} The simulated ES is intercepted and batched using a tumbling window approach (ensuring no overlap between windows). Predefined stream features are then extracted from each window and evaluated against the initial target feature values.
    \item \textbf{Hyperparameter Optimization (Bayesian Optimization):} A Bayesian optimization procedure iteratively adjusts the generator parameters based on the evaluation results from the previous step. This optimization loop continues through the generation and evaluation stages until the desired feature specifications are met or a maximum number of optimization attempts is reached, mirroring GEDI's formal optimization problem $ G(g_k) = \min_{\lambda \in \Lambda} d(f_e(A_\lambda), g_k) $.
\end{enumerate}
Upon conclusion of the optimization, \ourgeneratorname{} produces a static definition file compatible with DEF. This definition can be used to replay the stream with the optimized features. Furthermore, DEF's capability to combine multiple definitions allows for the simulation of concept drift in streams, enabling transitions (gradual or sudden) from one set of features to another.


\subsection{From Process Tree to Markov Chain}
\label{subsec:ptg_to_def}
The \ourgeneratorname{} pipeline starts with process trees, but its core simulation engine, DEF, operates on probabilistic Markov chain models. 
Thus, a critical step is converting the high-level process tree into a precise Markov chain representation. 
This transformation translates the tree's structural and behavioral properties into granular, probabilistic state transitions directly usable by DEF.

Our method adapts the approach by \cite{10904251} for converting an observed event log into a first-order Markov chain. 
To bridge the gap, \ourgeneratorname{} first generates a representative sample set of traces from the process tree using its inherent log generation capabilities (e.g., as provided by the PTLG framework \cite{Jouck2016PTandLogGeneratorAG}).
From this generated sample log \(L\), a mapping structure \(C\) records activity transition frequencies. For each trace, transitions (from an implicit start state \(s_{start}\), between consecutive activities \( (a_i, a_{i+1}) \), and to an implicit end state \(s_{end}\) are counted. These accumulated frequencies in \(C\) are then normalized to define the complete transition matrix of the Markov chain. This precisely constructed Markov chain is then supplied to DEF, enabling the simulation of ES whose probabilistic behavior accurately reflects the control-flow specified by the original process tree.

\subsection{Experimental Results}
\label{subsec:evaluation}
Similarly to \cite{maldonado_gedi_2024} we investigate two questions with our experiments: (E1)  How well does \ourgeneratorname{} produce ES with specific feature values? (feasibility) (E2) How well do the commonly used event logs represent the feature space covered by \ourgeneratorname{}? (representativeness)
Last, we also assess the utility of our generated ES and identify opportunities for further online streaming process mining algorithms developments.
Our generated data\footnote{\url{https://doi.org/10.5281/zenodo.16685512}} , implementation and evaluation results\footnote{\url{https://github.com/andreamalhera/gedi_streams/tree/icpm25}} are publicly available.

\begin{figure}
    \centering
    \includegraphics[width=0.7\linewidth]{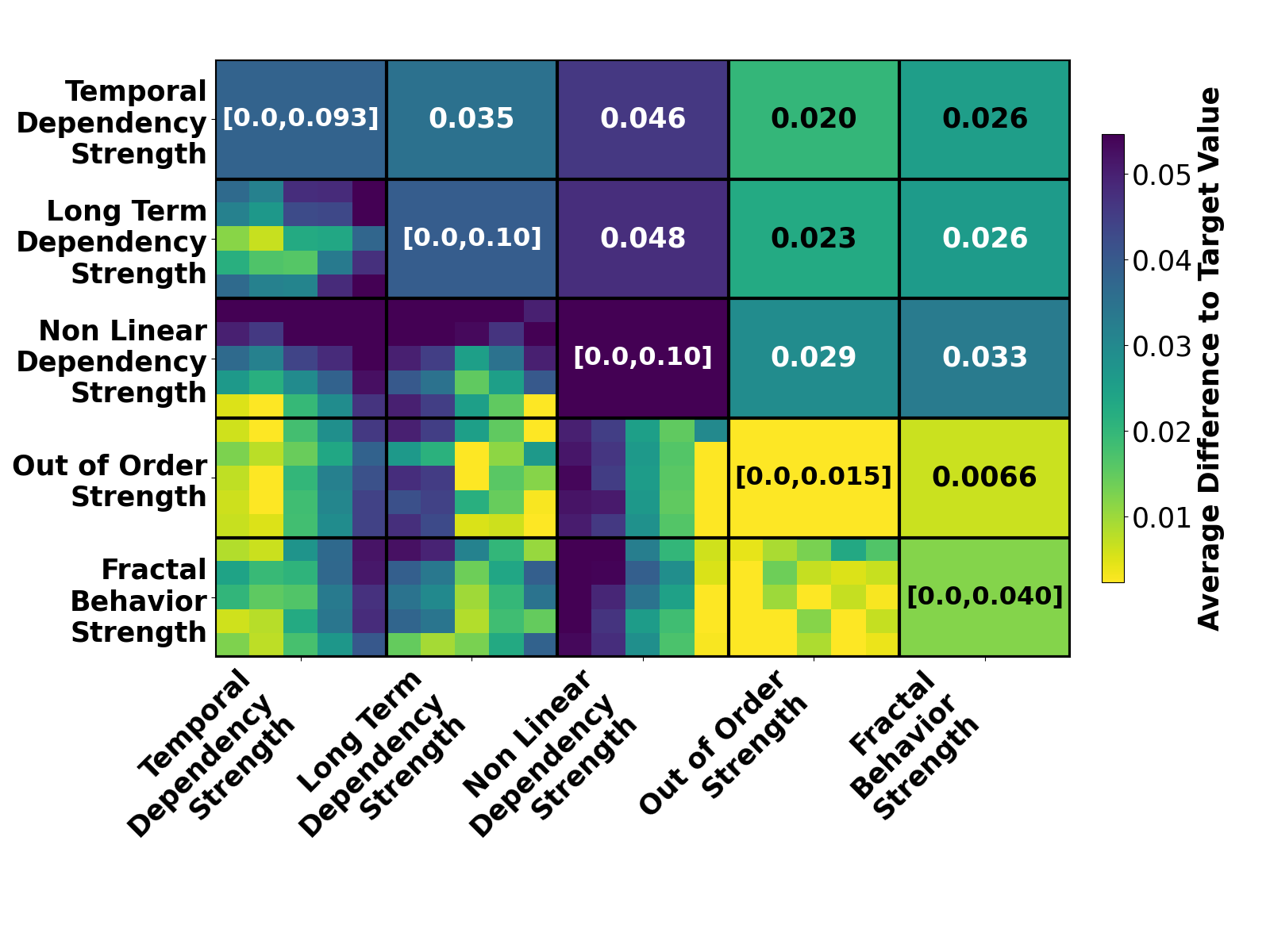}
    \caption{ In the diagonal, we present ranges with low distance results; above the diagonal, average distances; and below it, distances comparing targets to generated event stream feature values for each combination of features.}
    \label{fig:feasibility}
\end{figure}

We first evaluate the \textbf{feasibility} of ES generation by analyzing how closely our streams reach predefined target values (i.e., [0, 0.1, 0.5, 0.7, 1.0]).
We conducted 250 experiments covering all pairwise combinations of five features with five target values (i.e., $\binom{5}{2}=10 \;\to\; 5\cdot5=25 \;\to\; 10\times25=250$). Results are shown in \Cref{fig:feasibility}. 
Yellow indicates minimal deviation from the target, while dark purple indicates maximal.
Warmer colors denote easier-to-match features; cooler colors, harder ones. 
On the diagonal, we show feasible value ranges for individual features. 
The upper triangle demonstrates the average deviation for the pair of features. 
The lower triangle displays a grid $5\times5$ of target combinations.
For instance, \textit{Non Linear Dependency} combined with \textit{Temporal Dependency} or \textit{Long Term Dependency} is harder to match when trying to achieve higher target values for both. 
Similarly, \textit{Fractal Behavior} requires \textit{Non-Linear Dependencies}, given the high deviation when the target for \textit{Non Linear Dependencies} is set to zero. 
These results reveal the feasible regions for each feature and their pairwise interactions. This highlights constraints and dependencies that govern feasible targets in the ES generation space.

\begin{figure}
    \centering
    \includegraphics[width=0.8\linewidth]{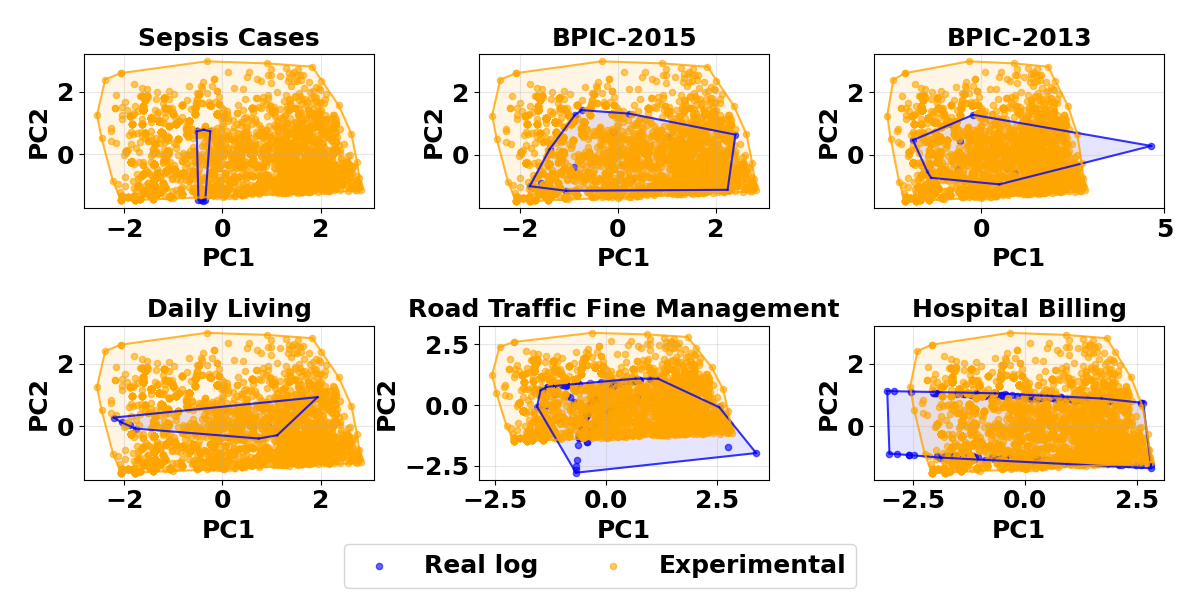}
    \caption{Principal-component map contrasting generated ES (orange) with benchmark event logs (blue).}
    \label{fig:representiveness}
\end{figure}

To answer our second evaluation question on
\textbf{representativeness}, we compare the 250 ES produced for (E1) with event logs commonly used to evaluate SPM algorithms. For that, we utilize the principal components (i.e., linear dimensionality reduction technique) to compare the achievable feature space of \ourgeneratorname{} with the aforementioned event logs. The results of this experiment can be found in \Cref{fig:representiveness}, where the orange hull represents the feature space from the ES and blue from the event log. The first observation is that no event log archives PC2 values higher than two. Some features are simply absent.

Thus, answering E2, investigated event logs do not contain most characteristics of realistic ES and thus undermine their applicability to evaluate SPM tasks.  


To further assess the \textbf{utility} of \ourgeneratorname{}, we conducted numerous experiments using established SPM algorithms 
\cite{burattin_uncovering_2023,burattin_online_2015,augusto_split_2019,burattin_heuristics_2012,burattin_control-flow_2014}.
from \cref{sec:intro}. 
Although these algorithms are valid and useful in multiple scenarios, their assumptions do not align with the characteristics defined in \cref{subsec:stream_features}. 
Given that, we encourage the development of novel streaming discovery approaches that can natively handle the complexities of realistic, concurrent ES. 

\section{Discussion and Conclusion}
\label{sec:discussion_and_conclusion}

In this paper, we address the gap between the SPM and data stream communities by identifying five key features from the literature that characterize realistic streams. As shown in \cref{sec:motivation_realistic_es}, including these features is crucial for real-time analysis, which existing methods often hinder through manual preprocessing. We expand the classical ES definition, propose a prototype generator, and show that \ourgeneratorname{} produces streams closely aligned with target features.

\textbf{Threads To Validity}: As no real-world streaming ES are publicly available, we could not compare our generated streams directly to such data. The selected features are reduced but cover a broad perspective in the literature \cite{9230157,burattin_uncovering_2023}. We also note a selection bias in the evaluated algorithms, as their assumptions differ and affect applicability.

\textbf{In future work}: We intend to further work on the intentional generation of event streams with controllable and explainable features selection. 

We would like to conclude by emphasizing that we do not intend to diminish the value of established methods for SPM, but rather to encourage the community to explore new directions by leveraging more realistic event streams, thereby enhancing algorithm robustness and applicability to real-world scenarios.

\paragraph{\textbf{Acknowledgments}}
This work received funding by the Deutsche Forschungsgemeinschaft (DFG), grant 496119880.

\bibliographystyle{splncs04}
\bibliography{references}

\end{document}